\documentclass[preprintnumbers,amsmath,amssymb,10pt,CJK]{revtex4}

\usepackage{graphicx}
\usepackage{dcolumn}
\usepackage{bm}

\begin{document}

\title{An  Implementation of a Positive Operator Valued
Measure}

\author{ Feng-Li Yan$^{1}$, Ting Gao$^{2}$, You-Cheng Li$^{1}$}

\affiliation {$^1$ College of Physics  and Information Engineering,
Hebei Normal
University, Shijiazhuang 050016, China\\
$^2$ College of Mathematics and Information Science, Hebei Normal
University, Shijiazhuang 050016, China}

\date{\today}

\begin{abstract}
An  implementation of the positive operator valued measure (POVM) is
given. By using this POVM one can realize the probabilistic
teleportation of an unknown two-particle state.
\end{abstract}
 \maketitle

 As a special case of the general measurement formalism, positive operator valued
measure (POVM)
 provides the simplest means by which one can study general
 measurement statistics, without the necessity for knowing the
 post-measurement state. They are viewed as a mathematical convenience that
 sometimes gives extra insight into quantum measurements \cite
 {NielsenChuang}.  Some of the time, one can perform a POVM to distinguish the states,
but never make  an error of mis-identification. Of course this
infallibility comes at the price that sometimes one obtains no
information about the identity of the state  \cite {NielsenChuang}.

Mor and Horodecki \cite {MorHorodecki} used POVM to the problems of
teleportation, which has been studied widely \cite {Bennett93, BPZ,
Boschi, Furusawa, GorbachevTrubilko, LuGuo, Bandyopadhyay, Rigolin,
DengLiLiZhouWang, DengLiLiZhouZhou, ShiGuo, ZengLong, LiuGuo, DaiLi,
YanTanYang, YanWang, YanYang, YanBai,
GaoYanWang,GaoWangYanChinPhysLett,GaoYanWangChinesePhysics,Gao2,
YDcpl} and has a number of useful applications in quantum
information and quantum computation \cite
{CiracZoller,Barenco,BennenttWiesner,Gisin,ShimizuImoto,Beige,BF,DengLong,
DengLong2,YanZhang,Gao,GaoYanWangJPA2,GaoYanWangIJMPC,GaoYanWangCPL3,MZL,ZMpra05,
ZYMLpra05,ZLMpra05,CaiLi}. It was shown that a perfect conclusive
teleportation can be obtained with any pure entangled states.
Recently, we proposed  a scheme  \cite
 {YDcpl} for probabilistic teleportation
 of an unknown two-particle state with a four-particle pure
 entangled state and the POVM. In Ref. \cite
 {YDcpl}   we  only gave  the POVM
 operators, but did not present the  concrete implementation of the POVM.
  In this Letter we will study this problem in details, especially an implementation of the POVM
 will be presented.

The optimal POVM  chosen by us in Ref. \cite {YDcpl} contains five
elements (here we use a  simple form, there is only a little
difference between them)
\begin{equation}
\begin{array}{ccc} P_i={q^2}|\Psi_i\rangle\langle \Psi_i|; &(i=1,2,3,4);&P_5=I-{q^2}\Sigma_{i=1}^4|\Psi_i\rangle\langle
\Psi_i|, \end{array}
\end{equation} where
\begin{equation}
\begin{array}{cc}|\Psi_1\rangle=\frac {1}{\alpha}|00\rangle+\frac {1}{\beta}|01\rangle+\frac {1}{\gamma}|10\rangle+\frac {1}{\delta}|11\rangle, &
|\Psi_2\rangle=\frac {1}{\alpha}|00\rangle+\frac {1}{\beta}|01\rangle-\frac {1}{\gamma}|10\rangle-\frac {1}{\delta}|11\rangle,\\
|\Psi_3\rangle=\frac {1}{\alpha}|00\rangle-\frac
{1}{\beta}|01\rangle+\frac {1}{\gamma}|10\rangle-\frac
{1}{\delta}|11\rangle,&
|\Psi_4\rangle=\frac {1}{\alpha}|00\rangle-\frac {1}{\beta}|01\rangle-\frac {1}{\gamma}|10\rangle+\frac {1}{\delta}|11\rangle,\\
\end{array}\end{equation}
$I$ is an identity operator, $\alpha, \beta, \gamma, \delta$ are
nonzero real numbers with $\frac {1}{\alpha^2}+\frac {1}{\beta^2} +
\frac {1}{\gamma^2}+\frac {1}{\delta^2}=1$, $q$ is a coefficient
relating to $\alpha,  \beta, \gamma,$ and $\delta$ such that $1\leq
 \frac {1}{q^2}\leq 4$, and makes $P_5$ to be a positive operator. Of
course if we choose $q^2=\frac {\mu^2}{4}$, where $\mu^2$ is the
smallest one in the set $\{\alpha^2, \beta^2, \gamma^2, \delta^2\}$,
we can obtain the highest probability of successful teleportation of
an unknown two-particle state \cite {YDcpl}.

In order to  implement this POVM,  we  construct a unitary operation
$U$, which satisfies
\begin{eqnarray}
U|\psi\rangle|000\rangle=\sqrt {P_1}|\psi\rangle|000\rangle+\sqrt
{P_2}|\psi\rangle|001\rangle +\sqrt
{P_3}|\psi\rangle|010\rangle+\sqrt
{P_4}|\psi\rangle|011\rangle+\sqrt {P_5}|\psi\rangle|100\rangle
\end{eqnarray}
for an  arbitrary  two-particle state $|\psi\rangle$.  Here the last
three qubits are  the auxiliary particles. When the unitary operator
 $U$ has been  performed, we can measure the auxiliary particles and
obtain the measurement result. It is easy to see that if $U$ is a
unitary operator such that
\begin{eqnarray}
U|00\rangle|000\rangle &&\nonumber=\frac {q}{\alpha}(\frac
{1}{\alpha}|00\rangle+\frac {1}{\beta}|01\rangle+\frac
{1}{\gamma}|10\rangle+\frac {1}{\delta}|11\rangle)|000\rangle +\frac
{q}{\alpha}(\frac {1}{\alpha}|00\rangle+\frac {1}{\beta}|01\rangle
-\frac {1}{\gamma}|10\rangle-\frac
{1}{\delta}|11\rangle)|001\rangle\\\nonumber
 &&+\frac
{q}{\alpha}(\frac {1}{\alpha}|00\rangle-\frac
{1}{\beta}|01\rangle+\frac {1}{\gamma}|10\rangle-\frac
{1}{\delta}|11\rangle)|010\rangle+\frac {q}{\alpha}(\frac
{1}{\alpha}|00\rangle-\frac {1}{\beta}|01\rangle-\frac
{1}{\gamma}|10\rangle+\frac
{1}{\delta}|11\rangle)|011\rangle+u|00\rangle|100\rangle,\\
\end{eqnarray}
\begin{eqnarray}
U|01\rangle|000\rangle &&\nonumber=\frac {q}{\beta}(\frac
{1}{\alpha}|00\rangle+\frac {1}{\beta}|01\rangle+\frac
{1}{\gamma}|10\rangle+\frac {1}{\delta}|11\rangle)|000\rangle+\frac
{q}{\beta}(\frac {1}{\alpha}|00\rangle +\frac
{1}{\beta}|01\rangle-\frac {1}{\gamma}|10\rangle-\frac
{1}{\delta}|11\rangle)|001\rangle\\\nonumber
 &&-\frac
{q}{\beta}(\frac {1}{\alpha}|00\rangle-\frac
{1}{\beta}|01\rangle+\frac {1}{\gamma}|10\rangle-\frac
{1}{\delta}|11\rangle)|010\rangle-\frac {q}{\beta}(\frac
{1}{\alpha}|00\rangle-\frac {1}{\beta}|01\rangle-\frac
{1}{\gamma}|10\rangle+\frac
{1}{\delta}|11\rangle)|011\rangle+v|01\rangle|100\rangle,\\
\end{eqnarray}
\begin{eqnarray}
U|10\rangle|000\rangle &&\nonumber=\frac {q}{\gamma}(\frac
{1}{\alpha}|00\rangle+\frac {1}{\beta}|01\rangle+ \frac
{1}{\gamma}|10\rangle+\frac {1}{\delta}|11\rangle)|000\rangle-\frac
{q}{\gamma}(\frac {1}{\alpha}|00\rangle +\frac
{1}{\beta}|01\rangle-\frac {1}{\gamma}|10\rangle-\frac
{1}{\delta}|11\rangle)|001\rangle\\\nonumber
 &&+\frac
{q}{\gamma}(\frac {1}{\alpha}|00\rangle-\frac
{1}{\beta}|01\rangle+\frac {1}{\gamma}|10\rangle-\frac
{1}{\delta}|11\rangle)|010\rangle -\frac {q}{\gamma}(\frac
{1}{\alpha}|00\rangle-\frac {1}{\beta}|01\rangle-\frac
{1}{\gamma}|10\rangle+\frac
{1}{\delta}|11\rangle)|011\rangle+w|10\rangle|100\rangle,\\
\end{eqnarray}
\begin{eqnarray}
U|11\rangle|000\rangle &&\nonumber=\frac {q}{\delta}(\frac
{1}{\alpha}|00\rangle+\frac {1}{\beta}|01\rangle+\frac {1}{\gamma}
|10\rangle+\frac {1}{\delta}|11\rangle)|000\rangle-\frac
{q}{\delta}(\frac {1}{\alpha}|00\rangle+\frac
{1}{\beta}|01\rangle-\frac {1}{\gamma}|10\rangle-\frac
{1}{\delta}|11\rangle)|001\rangle\\\nonumber
 &&-\frac
{q}{\delta}(\frac {1}{\alpha}|00\rangle-\frac
{1}{\beta}|01\rangle+\frac {1}{\gamma}|10\rangle-\frac
{1}{\delta}|11\rangle)|010\rangle +\frac {q}{\delta}(\frac
{1}{\alpha}|00\rangle-\frac {1}{\beta}|01\rangle-\frac
{1}{\gamma}|10\rangle+\frac
{1}{\delta}|11\rangle)|011\rangle+p|11\rangle|100\rangle,\\
\end{eqnarray}
then  Eq. (3) holds. Here
\begin{equation}
\begin{array}{cccc}u=\sqrt {1- \frac {4q^2}{\alpha^2}}, &v=\sqrt {1- \frac {4q^2}{\beta^2}},
&w=\sqrt {1- \frac {4q^2}{\gamma^2}}, &p=\sqrt {1- \frac {4q^2}{\delta^2}}.\\
\end{array}\end{equation}

For the sake of convenience we introduce the following parameters
\begin{equation}
\begin{array}{cccc}s=\sqrt {\alpha^2+\beta^2},&y=\sqrt
{\alpha^2+\delta^2},&z=\sqrt {\beta^2+\gamma^2},&t=\sqrt
{\gamma^2+\delta^2}. \end{array}\end{equation} A little thought
shows that  the following unitary operator {\begin{eqnarray} &&
~~~U=\nonumber\\
&&\left(\begin{array}{cccccccc}
\begin{array}{ccccc}
\frac {q}{\alpha^2}&\frac {\alpha}{2y}&\frac {\alpha}{2s} &\frac
{-\alpha}{2y}&\frac {u}{2\alpha}\\
 \frac {q}{\alpha^2}&\frac
{\alpha}{2y}&\frac {-\alpha}{2s}&\frac {\alpha}{2y}&\frac
{u}{2\alpha}\\
\frac {q}{\alpha^2}&\frac {-\alpha}{2y}&\frac {\alpha}{2s}&\frac {\alpha}{2y}&\frac {u}{2\alpha}\\
\frac {q}{\alpha^2}&\frac {-\alpha}{2y}&\frac {-\alpha}{2s}&\frac {-\alpha}{2y}&\frac {u}{2\alpha}\\
u&0&0&0&\frac {2q}{-\alpha}\\\end{array}&
~&\begin{array}{ccccc} \frac {q}{\alpha\beta}&\frac {\beta t}{2\gamma\delta s}&0&\frac {-\beta t}{2\gamma\delta s}&\frac {v}{2\alpha}\\
\frac {q}{\alpha\beta}&\frac {\beta t}{2\gamma\delta s}&0&\frac {\beta t}{2\gamma\delta s}&\frac {v}{2\alpha}\\
\frac {-q}{\alpha\beta}&\frac {\beta t}{2\gamma\delta s}&0&\frac {-\beta t}{2\gamma\delta s}&\frac {-v}{2\alpha}\\
\frac {-q}{\alpha\beta}&\frac {\beta t}{2\gamma\delta s}&0&\frac {\beta t}{2\gamma\delta s}&\frac {-v}{2\alpha}\\
0&0&0&0&0\\\end{array}&~&\begin{array}{ccccc}\frac {q}{\alpha\gamma}&\frac {\delta z}{2\beta\gamma y}&0&\frac {-\delta z}{2\beta\gamma y}&0\\
\frac {-q}{\alpha\gamma}&\frac {-\delta z}{2\beta\gamma y}&0&\frac {-\delta z}{2\beta\gamma y}&0\\
\frac {q}{\alpha\gamma}&\frac {-\delta z}{2\beta\gamma y}&0&\frac {\delta z}{2\beta\gamma y}&0\\
\frac {-q}{\alpha\gamma}&\frac {\delta z}{2\beta\gamma y}&0&\frac {\delta z}{2\beta\gamma y}&0\\
0&0&0&0&0\\\end{array}&~&
\begin{array}{ccccc}
\frac {q}{\alpha\delta}&\frac {w}{2\alpha}&0&\frac {\alpha}{2s}&\frac {p}{2\alpha}\\
\frac {-q}{\alpha\delta}&\frac {-w}{2\alpha}&0&\frac {\alpha}{2s}&\frac {-p}{2\alpha}\\
\frac {-q}{\alpha\delta}&\frac {w}{2\alpha}&0&\frac {\alpha}{2s}&\frac {-p}{2\alpha}\\
\frac {q}{\alpha\delta}&\frac {-w}{2\alpha}&0&\frac {\alpha}{2s}&\frac {p}{2\alpha}\\
0&0&0&0&0\\\end{array}\\
~&I_3&~&~&~&~&~&~\\
\begin{array}{ccccc}
\frac {q}{\alpha\beta}&0&\frac {-\beta}{2s}&0&\frac {u}{2\beta}\\
\frac {q}{\alpha\beta}&0&\frac {\beta}{2s}&0&\frac {u}{2\beta}\\
\frac {-q}{\alpha\beta}&0&\frac {\beta}{2s}&0&\frac {-u}{2\beta}\\
\frac {-q}{\alpha\beta}&0&\frac {-\beta}{2s}&0&\frac {-u}{2\beta}\\
0&0&0&0&0\\
\end{array}&~&\begin{array}{ccccc}
\frac {q}{\beta^2}&\frac {\alpha t}{2\gamma\delta s}&\frac {\beta}{2z}&\frac {-\alpha t}{2\gamma\delta s}&\frac {v}{2\beta}\\
\frac {q}{\beta^2}&\frac {\alpha t}{2\gamma\delta s}&\frac {-\beta}{2z}&\frac {\alpha t}{2\gamma\delta s}&\frac {v}{2\beta}\\
\frac {q}{\beta^2}&\frac {-\alpha t}{2\gamma\delta s}&\frac {\beta}{2z}&\frac {\alpha t}{2\gamma\delta s}&\frac {v}{2\beta}\\
\frac {q}{\beta^2}&\frac {-\alpha t}{2\gamma\delta s}&\frac {-\beta}{2z}&\frac {-\alpha t}{2\gamma\delta s}&\frac {v}{2\beta}\\
v&0&0&0&\frac {2q}{-\beta}\\
\end{array}&~&
\begin{array}{ccccc}
\frac {q}{\beta\gamma}&\frac {-\gamma y}{2\alpha\delta z}&0&\frac {\gamma y}{2\alpha\delta z}&0\\
\frac {-q}{\beta\gamma}&\frac {\gamma y}{2\alpha\delta z}&0&\frac {\gamma y}{2\alpha\delta z}&0\\
\frac {-q}{\beta\gamma}&\frac {-\gamma y}{2\alpha\delta z}&0&\frac {\gamma y}{2\alpha\delta z}&0\\
\frac {q}{\beta\gamma}&\frac {\gamma y}{2\alpha\delta z}&0&\frac {\gamma y}{2\alpha\delta z}&0\\
0&0&0&0&0\\
\end{array}&~&\begin{array}{ccccc}
\frac {q}{\beta\delta}&\frac {w}{2\beta}&\frac {\beta}{2z}&\frac {-\beta}{2s}&\frac {p}{2\beta}\\
\frac {-q}{\beta\delta}&\frac {-w}{2\beta}&\frac {\beta}{2z}&\frac {-\beta}{2s}&\frac {-p}{2\beta}\\
\frac {q}{\beta\delta}&\frac {-w}{2\beta}&\frac {\beta}{2z}&\frac {\beta}{2s}&\frac {p}{2\beta}\\
\frac {-q}{\beta\delta}&\frac {w}{2\beta}&\frac {\beta}{2z}&\frac {\beta}{2s}&\frac {-p}{2\beta}\\
0&0&0&0&0\\\end{array}&~\\
~&~&~&I_3&~&~&~&~\\
\begin{array}{ccccc}
\frac {q}{\alpha\gamma}&0&0&0&\frac {u}{2\gamma}\\
\frac {-q}{\alpha\gamma}&0&0&0&\frac {-u}{2\gamma}\\
\frac {q}{\alpha\gamma}&0&0&0&\frac {u}{2\gamma}\\
\frac {-q}{\alpha\gamma}&0&0&0&\frac {-u}{2\gamma}\\
0&0&0&0&0\\
\end{array}&~&\begin{array}{ccccc}
\frac {q}{\beta\gamma}&\frac {-\delta s}{2\alpha\beta t}&\frac {-\gamma}{2z}&\frac {\delta s}{2\alpha\beta t}&\frac {v}{2\gamma}\\
\frac {-q}{\beta\gamma}&\frac {\delta s}{2\alpha\beta t}&\frac {-\gamma}{2z}&\frac {\delta s}{2\alpha\beta t}&\frac {-v}{2\gamma}\\
\frac {-q}{\beta\gamma}&\frac {-\delta s}{2\alpha\beta t}&\frac {\gamma}{2z}&\frac {\delta s}{2\alpha\beta t}&\frac {-v}{2\gamma}\\
\frac {q}{\beta\gamma}&\frac {\delta s}{2\alpha\beta t}&\frac {\gamma}{2z}&\frac {\delta s}{2\alpha\beta t}&\frac {v}{2\gamma}\\
0&0&0&0&0\\
\end{array}&~&
\begin{array}{ccccc}
\frac {q}{\gamma^2}&\frac {-\beta y}{2\alpha\delta z}&\frac {\gamma}{2t}&\frac {\beta y}{2\alpha\delta z}&\frac {\gamma}{2t}\\
\frac {q}{\gamma^2}&\frac {-\beta y}{2\alpha\delta z}&\frac {-\gamma}{2t}&\frac {-\beta y}{2\alpha\delta z}&\frac {\gamma}{2t}\\
\frac {q}{\gamma^2}&\frac {\beta y}{2\alpha\delta z}&\frac {\gamma}{2t}&\frac {-\beta y}{2\alpha\delta z}&\frac {\gamma}{2t}\\
\frac {q}{\gamma^2}&\frac {\beta y}{2\alpha\delta z}&\frac {-\gamma}{2t}&\frac {\beta y}{2\alpha\delta z}&\frac {\gamma}{2t}\\
w&0&0&0&0\\
\end{array}&~&
\begin{array}{ccccc}
\frac {q}{\gamma\delta}&\frac {w}{2\gamma}&\frac {-\gamma}{2z}&0&\frac {p}{2\gamma}\\
\frac {q}{\gamma\delta}&\frac {w}{2\gamma}&\frac {\gamma}{2z}&0&\frac {p}{2\gamma}\\
\frac {-q}{\gamma\delta}&\frac {w}{2\gamma}&\frac {\gamma}{2z}&0&\frac {-p}{2\gamma}\\
\frac {-q}{\gamma\delta}&\frac {w}{2\gamma}&\frac {-\gamma}{2z}&0&\frac {-p}{2\gamma}\\
0&\frac {2q}{-\gamma}&0&0&0\\
\end{array}&~\\
~&~&~&~&~&I_3&~&~\\
\begin{array}{ccccc}
\frac {q}{\alpha\delta}&\frac {-\delta}{2y}&0&\frac {\delta}{2y}&\frac {u}{2\delta}\\
\frac {-q}{\alpha\delta}&\frac {\delta}{2y}&0&\frac {\delta}{2y}&\frac {-u}{2\delta}\\
\frac {-q}{\alpha\delta}&\frac {-\delta}{2y}&0&\frac {\delta}{2y}&\frac {-u}{2\delta}\\
\frac {q}{\alpha\delta}&\frac {\delta}{2y}&0&\frac {\delta}{2y}&\frac {u}{2\delta}\\
0&0&0&0&0\\
\end{array}&~&
\begin{array}{ccccc}
\frac {q}{\beta\delta}&\frac {-\gamma s}{2\alpha\beta t}&0&\frac {\gamma s}{2\alpha\beta t}&\frac {v}{2\delta}\\
\frac {-q}{\beta\delta}&\frac {\gamma s}{2\alpha\beta t}&0&\frac {\gamma s}{2\alpha\beta t}&\frac {-v}{2\delta}\\
\frac {q}{\beta\delta}&\frac {\gamma s}{2\alpha\beta t}&0&\frac {-\gamma s}{2\alpha\beta t}&\frac {v}{2\delta}\\
\frac {-q}{\beta\delta}&\frac {-\gamma s}{2\alpha\beta t}&0&\frac {-\gamma s}{2\alpha\beta t}&\frac {-v}{2\delta}\\
0&0&0&0&0\\
\end{array}&~&
\begin{array}{ccccc}
\frac {q}{\gamma\delta}&\frac {\alpha z}{2\beta\gamma y}&\frac {-\delta}{2t}&\frac {-\alpha z}{2\beta\gamma y}&\frac {-\delta}{2t}\\
\frac {q}{\gamma\delta}&\frac {\alpha z}{2\beta\gamma y}&\frac {\delta}{2t}&\frac {\alpha z}{2\beta\gamma y}&\frac {-\delta}{2t}\\
\frac {-q}{\gamma\delta}&\frac {\alpha z}{2\beta\gamma y}&\frac {\delta}{2t}&\frac {-\alpha z}{2\beta\gamma y}&\frac {\delta}{2t}\\
\frac {-q}{\gamma\delta}&\frac {\alpha z}{2\beta\gamma y}&\frac {-\delta}{2t}&\frac {\alpha z}{2\beta\gamma y}&\frac {\delta}{2t}\\
0&0&0&0&0\\
\end{array}&~&
\begin{array}{ccccc}
\frac {q}{\delta^2}&\frac {w}{2\delta}&0&0&\frac {p}{2\delta}\\
\frac {q}{\delta^2}&\frac {w}{2\delta}&0&0&\frac {p}{2\delta}\\
\frac {q}{\delta^2}&\frac {-w}{2\delta}&0&0&\frac {p}{2\delta}\\
\frac {q}{\delta^2}&\frac {-w}{2\delta}&0&0&\frac {p}{2\delta}\\
p&0&0&0&\frac {2q}{-\delta}\\
\end{array}&~\\
~&~&~&~&~&~&~&I_3\\
\end{array}\right)\nonumber\\
 \end{eqnarray}} can satisfy
Eq.(4)-(7). Here $I_3$ is a $3\times 3$ identity matrix.

We can  prove that $U$ can be expressed as
\begin{eqnarray}\nonumber
U=&&M\left (\begin{array}{cc}
\frac {\alpha}{s}&\frac {-\beta}{s}\\
\frac {-\beta}{s}&\frac {-\alpha}{s}\\
\end{array}\right )_{1, 10}
\left (\begin{array}{cc}
\frac {\alpha}{y}&\frac {-\delta}{y}\\
\frac {-\delta}{y}&\frac {-\alpha}{y}\\
\end{array}\right )_{2, 27}\left (\begin{array}{cc}
\frac {\alpha}{s}&\frac {\beta}{s}\\
\frac {\beta}{s}&\frac {-\alpha}{s}\\
\end{array}\right )_{3, 12}\left (\begin{array}{cc}
\frac {\alpha}{y}&\frac {\delta}{y}\\
\frac {\delta}{y}&\frac {-\alpha}{y}\\
\end{array}\right )_{4, 25}\\\nonumber
&&\left (\begin{array}{cc}
\frac {\beta}{z}&\frac {\gamma}{z}\\
\frac {\gamma}{z}&\frac {-\beta}{z}\\
\end{array}\right )_{9, 20}\left (\begin{array}{cc}
\frac {\beta}{z}&\frac {-\gamma}{z}\\
\frac {-\gamma}{z}&\frac {-\beta}{z}\\
\end{array}\right )_{11, 18}\left (\begin{array}{cc}
\frac {\gamma}{t}&\frac {-\delta}{t}\\
\frac {-\delta}{t}&\frac {-\gamma}{t}\\
\end{array}\right )_{17, 26}
\left (\begin{array}{cc}
\frac {\gamma}{t}&\frac {\delta}{t}\\
\frac {\delta}{t}&\frac {-\gamma}{t}\\
\end{array}\right )_{19, 28}\\\nonumber
&&\left (\begin{array}{cc}
\frac {-t}{\gamma\delta}&\frac {-s}{\alpha\beta}\\
\frac {-s}{\alpha\beta}&\frac {t}{\gamma\delta}\\
\end{array}\right )_{10, 28}
\left (\begin{array}{cc}
\frac {-t}{\gamma\delta}&\frac {-s}{\alpha\beta}\\
\frac {-s}{\alpha\beta}&\frac {t}{\gamma\delta}\\
\end{array}\right )_{12, 26}
\left (\begin{array}{cc}
\frac {y}{\alpha\delta}&\frac {-z}{\beta\gamma}\\
\frac {-z}{\beta\gamma}&\frac {-y}{\alpha\delta}\\
\end{array}\right )_{18, 25}
\left (\begin{array}{cc}
\frac {y}{\alpha\delta}&\frac {z}{\beta\gamma}\\
\frac {z}{\beta\gamma}&\frac {-y}{\alpha\delta}\\
\end{array}\right )_{20, 27}\\\nonumber
&&\left (\begin{array}{cc}
\frac {-2q}{\alpha}&u\\
u&\frac {2q}{\alpha}\\
\end{array}\right )_{5, 28}
\left (\begin{array}{cc}
\frac {-2q}{\beta}&v\\
v&\frac {2q}{\beta}\\
\end{array}\right )_{13, 27}
\left (\begin{array}{cc}
w&\frac {-2q}{\gamma}\\
\frac {-2q}{\gamma}&-w\\
\end{array}\right )_{21, 26}
\left (\begin{array}{cc}
\frac {2q}{\delta}&p\\
p&\frac {-2q}{\delta}\\
\end{array}\right )_{25, 29}\\\nonumber
&&\left (\begin{array}{cc}
0&1\\
1&0\\
\end{array}\right )_{1, 28}\left (\begin{array}{cc}
0&1\\
1&0\\
\end{array}\right )_{9, 27}\left (\begin{array}{cc}
0&1\\
1&0\\
\end{array}\right )_{17, 21},\\
\end{eqnarray}
where
\begin{eqnarray}\nonumber
M=&&\left(\begin{array}{cc} \frac {1}{\sqrt 2}&
\frac {1}{\sqrt 2}\\
\frac {1}{\sqrt 2}& \frac {-1}{\sqrt 2}\\
\end{array}\right)_{1,2}
\left(\begin{array}{cc} \sqrt{\frac {2}{3}}&
\frac {1}{\sqrt 3}\\
\frac {1}{\sqrt 3}& -\sqrt{\frac {2}{3}}\\
\end{array}\right)_{1,3}\left(\begin{array}{cc}\frac {\sqrt{3}}{2}&
\frac {1}{2}\\
\frac {1}{2}& -\frac {\sqrt{3}}{2}\\
\end{array}\right)_{1,4}\left(\begin{array}{cc}0&
1\\
1& 0\\
\end{array}\right)_{2,4}
\left(\begin{array}{cc}\sqrt {\frac {1}{3}}&
\frac {\sqrt 2}{\sqrt 3}\\
\frac {\sqrt 2}{\sqrt 3}& -\sqrt {\frac {1}{3}}\\
\end{array}\right)_{2,3}
\left(\begin{array}{cc} {\frac {1}{\sqrt 2}}&
\frac {1}{\sqrt 2}\\
\frac {1}{\sqrt 2}& -\frac {1}{\sqrt 2}\\
\end{array}\right)_{3,4}\\\nonumber
~~&&\left(\begin{array}{cc} \frac {1}{\sqrt 2}&
\frac {1}{\sqrt 2}\\
\frac {1}{\sqrt 2}& \frac {-1}{\sqrt 2}\\
\end{array}\right)_{9,10}
\left(\begin{array}{cc} \sqrt{\frac {2}{3}}&
\frac {1}{\sqrt 3}\\
\frac {1}{\sqrt 3}& -\sqrt{\frac {2}{3}}\\
\end{array}\right)_{9,11}\left(\begin{array}{cc}\frac {\sqrt{3}}{2}&
\frac {1}{2}\\
\frac {1}{2}& -\frac {\sqrt{3}}{2}\\
\end{array}\right)_{9,12}\left(\begin{array}{cc}0&
1\\
1& 0\\
\end{array}\right)_{10,12}
\left(\begin{array}{cc}\sqrt {\frac {1}{3}}&
\frac {\sqrt 2}{\sqrt 3}\\
\frac {\sqrt 2}{\sqrt 3}& -\sqrt {\frac {1}{3}}\\
\end{array}\right)_{10,11}
\left(\begin{array}{cc} {\frac {1}{\sqrt 2}}&
\frac {1}{\sqrt 2}\\
\frac {1}{\sqrt 2}& -\frac {1}{\sqrt 2}\\
\end{array}\right)_{11,12}\\\nonumber
~~&&\left(\begin{array}{cc} \frac {1}{\sqrt 2}&
\frac {1}{\sqrt 2}\\
\frac {1}{\sqrt 2}& \frac {-1}{\sqrt 2}\\
\end{array}\right)_{17,18}
\left(\begin{array}{cc} \sqrt{\frac {2}{3}}&
\frac {1}{\sqrt 3}\\
\frac {1}{\sqrt 3}& -\sqrt{\frac {2}{3}}\\
\end{array}\right)_{17,19}\left(\begin{array}{cc}\frac {\sqrt{3}}{2}&
\frac {1}{2}\\
\frac {1}{2}& -\frac {\sqrt{3}}{2}\\
\end{array}\right)_{17,20}\left(\begin{array}{cc}0&
1\\
1& 0\\
\end{array}\right)_{18,20}
\left(\begin{array}{cc}\sqrt {\frac {1}{3}}&
\frac {\sqrt 2}{\sqrt 3}\\
\frac {\sqrt 2}{\sqrt 3}& -\sqrt {\frac {1}{3}}\\
\end{array}\right)_{18,19}
\left(\begin{array}{cc} {\frac {1}{\sqrt 2}}&
\frac {1}{\sqrt 2}\\
\frac {1}{\sqrt 2}& -\frac {1}{\sqrt 2}\\
\end{array}\right)_{19,20}\\\nonumber
~~&&\left(\begin{array}{cc} \frac {1}{\sqrt 2}&
\frac {1}{\sqrt 2}\\
\frac {1}{\sqrt 2}& \frac {-1}{\sqrt 2}\\
\end{array}\right)_{25,26}
\left(\begin{array}{cc} \sqrt{\frac {2}{3}}&
\frac {1}{\sqrt 3}\\
\frac {1}{\sqrt 3}& -\sqrt{\frac {2}{3}}\\
\end{array}\right)_{25,27}\left(\begin{array}{cc}\frac {\sqrt{3}}{2}&
\frac {1}{2}\\
\frac {1}{2}& -\frac {\sqrt{3}}{2}\\
\end{array}\right)_{25,28}\left(\begin{array}{cc}0&
1\\
1& 0\\
\end{array}\right)_{26,28}
\left(\begin{array}{cc}\sqrt {\frac {1}{3}}&
\frac {\sqrt 2}{\sqrt 3}\\
\frac {\sqrt 2}{\sqrt 3}& -\sqrt {\frac {1}{3}}\\
\end{array}\right)_{26,27}
\left(\begin{array}{cc} {\frac {1}{\sqrt 2}}&
\frac {1}{\sqrt 2}\\
\frac {1}{\sqrt 2}& -\frac {1}{\sqrt 2}\\
\end{array}\right)_{27,28}.\\
\end{eqnarray}
Here  $\Omega=\left(\begin{array}{cc}
\xi & \zeta\\
\zeta & -\xi\\
\end{array}\right)_{i,j}
$ denotes a $32\times 32$  unitary matrix or a two-level unitary
matrix, where the matrix elements $\Omega_{i,i}=\xi,
\Omega_{i,j}=\Omega_{j,i}=\zeta, \Omega_{j,j}=-\xi,$
$\Omega_{l,l}=1,$ for $l\neq i,j$, and the other elements of matrix
$\Omega$ are zero.

By making use of Gray codes  one can construct a circuit
implementing a two-level unitary operator $\Omega$, where the
circuit only  consists of a number of controlled operations \cite
{NielsenChuang}.

Barenco et al exhibited  a general simulation of controlled
operations using only one-bit gates  and the two-bit controlled-not
(CNOT) gates \cite {Barenco2}. Combining the results obtained by
Barenco et al. and the above decomposition of $U$ we can give the
explicit construction of the unitary operation $U$ using one-qubit
gates and two-qubit CNOT gates. For save space, we do not give an
implementation of $U$ in terms of one and two qubit operations and
also do not depict out the quantum circuit illustrating the
procedure of the implementation of the POVM.

In summary, an implementation of a POVM has been given. By using
this POVM one can realize the probabilistic teleportation of an
unknown two-particle state. We hope that this
 POVM will be realized by experiment, furthermore
 one will really see the  probabilistic teleportation
 of an unknown two-particle state with a four-particle pure
 entangled state and POVM.\\[0.2cm]

{\noindent\bf Acknowledgments}\\[0.2cm]

 This work was supported by the National  Natural Science Foundation of
China under Grant No: 10671054 and Hebei Natural Science Foundation
of China under Grant No: A2005000140.

\end{document}